\documentclass[11pt]{article}
\pdfoutput=1
\usepackage{jcappub}
\usepackage{slashed}
\usepackage{stackengine}
\usepackage{amsmath}
\usepackage[frak=euler,scr=boondox,bb= pazo]{mathalfa}

\def\be{\begin{equation}}
\def\ee{\end{equation}}
\def\bea{\begin{eqnarray}}
\def\eea{\end{eqnarray}}

\begin{document}

\title{The world as a neural network} 

\author{Vitaly Vanchurin}

\emailAdd{vvanchur@d.umn.edu}

\date{\today}

\affiliation{Department of Physics, University of Minnesota, Duluth, Minnesota, 55812 \\
Duluth Institute for Advanced Study, Duluth, Minnesota, 55804}

\abstract{

We discuss a possibility that the entire universe on its most fundamental level is a neural network. We identify two different types of dynamical degrees of freedom: ``trainable'' variables  (e.g. bias  vector or weight matrix) and ``hidden'' variables (e.g. state vector of neurons). We first consider stochastic evolution of the trainable variables to argue that near equilibrium their dynamics is well approximated by Madelung equations (with free energy representing the phase) and further away from the equilibrium by Hamilton-Jacobi equations (with free energy representing the Hamilton's principal function). This shows that the trainable variables can indeed exhibit classical and quantum behaviors with the state vector of neurons representing the hidden variables. We then study stochastic evolution of the hidden variables by considering $D$ non-interacting subsystems with average state vectors, $\bar{\bf x}^{1}$, ..., $\bar{\bf x}^{D}$ and an overall average state vector $\bar{\bf x}^{0}$. In the limit when the weight matrix is a permutation matrix,  the dynamics of $\bar{\bf x}^{\mu}$ can be described in terms of relativistic strings in an emergent $D+1$ dimensional Minkowski space-time. If the subsystems are minimally interacting, with interactions described by a metric tensor, then the emergent space-time becomes curved. We argue that the entropy production in such a system is a local function of the metric tensor which should be determined by the symmetries of the Onsager tensor. It turns out that a very simple and highly symmetric Onsager tensor leads to the entropy production described by the Einstein-Hilbert term. This shows that the learning dynamics of a neural network can indeed exhibit approximate behaviors described by both quantum mechanics and general relativity. We also discuss a possibility that the two descriptions are holographic duals of each other.
 }

\maketitle

\section{Introduction}

Quantum mechanics is a remarkably successful paradigm for modeling physical phenomena on a wide range of scales ranging from  $10^{-19}$ meters (i.e. high-energy experiments) to $10^{+26}$ meters (i.e. cosmological observations.) The paradigm is so successful that it is widely believed that on the most fundamental level the entire universe is governed by the rules of quantum mechanics and even gravity should somehow emerge from it. This is known as the problem of quantum gravity that so far has not been solved, but some progress had been made in context of AdS/CFT \cite{Maldacena, Witten, Susskind}, loop quantum gravity \cite{Ashtekar, Smolin, ABL} and emergent gravity \cite{Jacobson, Verlinde, Padmanabhan}.  Although extremely important, the problem of quantum gravity is not the only problem with quantum mechanics. The quantum framework also starts to fall apart with introduction of observers. Everything seems to work very well when observers are kept outside of a quantum system, but it is far less clear how to describe macroscopic observers in a quantum system such as the universe itself. The realization of the problem triggered an ongoing debate on the interpretations of quantum mechanics, which remains unsettled to this day. On one side of the debate, there  are proponents of the many-worlds interpretation claiming that everything in the universe (including observers) must be governed by the Schr\"odinger equation \cite{Everett}, but then it is not clear how classical probabilities would emerge. One the other side of the debate, there are proponents of the hidden variables theories \cite{Bohm}, but there it is also unclear what is the role of the complex wave-function in a purely statistical system. It is important to emphasize that a working definition of observers is necessary not only for settling some philosophical debates, but for understanding the results of real physical experiments and cosmological observations. In particular, a self-consistent and paradoxes-free definition of observers would allow us to understand the significance of  Bell's inequalities \cite{Bell} and to make probabilistic prediction in cosmology \cite{Measure}. To resolve the apparent inconsistency  (or incompleteness) in our description of the physical world, we shall entertain an idea of having a more fundamental theory than quantum mechanics. A working hypothesis is that on the most fundamental level the dynamics of the entire universe is described by a microscopic neural network which undergoes learning evolution. If correct, then not only macroscopic observers but, more importantly, quantum mechanics and general relativity should correctly describe the dynamics of the microscopic neural network in the appropriate limits.\footnote{The idea of using neural networks to describe gravity was recently explored in Ref. \cite{Dvali} in context of quantum neural networks, in Ref. \cite{Hashimoto} in context of AdS/CFT and in Ref. \cite{machine_learning} in context of emergent gravity.} 

In this paper we shall first demonstrate that near equilibrium the learning evolution of a neural network can indeed be modeled (or approximated) with the Madelung equations (see Sec. \ref{sec:Quantum}), where the phase of the complex wave-function has a precise physical interpretation as the free energy of a statistical ensemble of hidden variables. The hidden variables describe the state of the individual neurons whose statistical ensemble is given by a partition function and the corresponding free energy. This free energy is a function of the trainable variables (such as bias vector and weight matrix)  whose  stochastic and learning dynamics we shall study (see Sec. \ref{sec:Entropic}). Note that while the stochastic dynamics generically leads to the production of entropy (i.e. second law of thermodynamics) the learning dynamics generically leads to the destruction of entropy (i.e. second law of learning). As a result in the equilibrium  the time-averaged entropy of the system remains constant and the corresponding dynamics can be modeled using quantum mechanics. It is important to note that the entropy (and entropy production) that we discuss here is the entropy of either hidden or trainable variables which need not vanish even for pure states. Of course, one can also discuss mixed states and then the corresponding von Neumann entropy gives an additional contribution to the total entropy.

The situation changes dramatically, whenever some of the degrees of freedom are not thermalized. While it should in principle be possible to model the thermalized degrees of freedom using quantum theory, the non-thermalized degrees of freedom are not likely to follow exactly the rules of quantum mechanics. We shall discuss two non-equilibrium limits: one which can nevertheless be described using  classical physics (e.g. Hamiltonian mechanics) and the other one which can be described using gravitational physics (e.g general relativity). The classical limit is relevant when the non-equilibrium evolution of the trainable variables is dominated by the entropy destruction due to learning, but the stochastic entropy production is negligible. The dynamics of such a system is well approximated by the Hamilton-Jacobi equations with free energy playing the role of the Hamilton's principal function (see Sec. \ref{sec:Hamiltonian}). The gravitational limit is relevant when even the hidden variables (i.e. state vectors of neurons) have not yet thermalized and the stochastic entropy production governs the non-equilibrium evolution of the system (see Sec. \ref{sec:Gravity}).  In the long run all of the degrees of freedom must thermalize and then quantum mechanics should provide a correct description of the learning system.  

It is well known that during learning the neural network is attracted towards a network with a low complexity, a phenomenon also known as dimensional reduction or what we call the second law of learning \cite{machine_learning}.  An example of a low complexity neural network is the one described by a permutation weight matrix or when the neural network is made out of one-dimensional chains of neurons.\footnote{A similar phenomenon was recently observed in context of the information graph flow \cite{graphflow}.} If the set of state vectors can also be divided into non-interacting subsets (or subsystems) with average state vectors, $\bar{\bf x}^{1}$, ..., $\bar{\bf x}^{D}$ and an overall average state vector $\bar{\bf x}^{0}$, then the dynamics of $\bar{\bf x}^\mu$ can be described with relativistic strings in an emergent $D+1$ dimensional space-time (see Sec. \ref{sec:Strings}). In general, the subsystems would interact and then the emergent space-time would be described by a gravitational theory such as general relativity (see Sec. \ref{sec:Gravity}).  Note that, in either case, the main challenge is to figure out exactly which degrees of freedom have already thermalized (and thus can be modeled with quantum mechanics) and which degrees of freedom are still in the process of thermalization and should be modeled with other methods such as Hamiltonian mechanics or general relativity. In addition, we shall discuss yet another method which is motivated by the holographic principle and is particularly useful when the bulk neurons are still in the process of equilibration,  but the boundary neurons have already thermalized (see Sec. \ref{sec:Holography}).

The paper is organized as follows.  In Sec. \ref{sec:Review} we review the theory of neural networks and in Sec. \ref{sec:Thermodynamics} we discuss a thermodynamic approach to learning. In Sec. \ref{sec:Entropic} we derive the action which governs dynamics of the trainable variables by applying the principle of stationary entropy production.  The action is used to study the dynamics near equilibrium in Secs. \ref{sec:Quantum} (which corresponds to quantum limit) and further away from equilibrium in Sec. \ref{sec:Hamiltonian} (which corresponds to classical limit). In Sec. \ref{sec:Hidden} we study a non-equilibrium dynamics of the hidden variables and in Sec. \ref{sec:Strings} we argue that in certain limits the dynamics can be described in terms of relativistic strings in the emergent space-time. In Sec. \ref{sec:Gravity} we apply the principle of stationary entropy production to derive the action which describes equilibration of the emergent space-time (which corresponds to gravitational limit) and in Sec. \ref{sec:Holography} we discuss when the gravitational theory can have a holographic dual description as a quantum theory. In Sec. \ref{sec:Discussion} we summarize and discuss the main results of the paper.

\section{Neural networks}\label{sec:Review}

We start with a brief review of the theory of neural networks by following the construction that was introduced in Ref. \cite{machine_learning}. The neural network shall be defined as a neural septuple $({\bf x}, \hat{P}_{in}, \hat{P}_{out}, \hat{w}, {\bf b}, {f}, H)$, where ${\bf x} \in  \mathbb{R}^N $,  is the state vector of neurons,  $\hat{P}_{in} $ and  $\hat{P}_{out}$ are the projection operators to subspaces spanned by respectively, $N_{in}$, input and, $N_{out}$, output neurons,  $\hat{w} \in \mathbb{R}^{N\times N} $, is a weight matrix, ${\bf b} \in \mathbb{R}^N$ is a bias vector,  ${f}: \mathbb{R} \rightarrow \mathbb{R} $ is an activation function and  ${H}:    \mathbb{R}^N \times \mathbb{R}^N  \times \mathbb{R}^{N\times N} \rightarrow \mathbb{R}$ is a loss function. This definition is somewhat different from the one usually used in the literature on machine learning, but we found that it is a lot more useful for analyzing physical theories in context of a microscopic neural network that we are interested in here. We shall not distinguish between different layers and so all $N$ neurons are connected into a single neural network with connections described by a single $N\times N$ weight matrix, $\hat{w}$. The matrix can be viewed as an adjacency matrix of a weighted directed graph with neurons representing the nodes and elements of the weight matrix representing directed edges. However, we will distinguish between two different types of neurons: the boundary neurons, $N_{\partial}=N_{in}+N_{out}$,  and the bulk neurons, $N_{\slashed{\partial}} = N-N_{\partial}$. Similarly, the boundary and the bulk projection operators are defined respectively as $ \hat{P}_\partial=\hat{P}_{in}+\hat{P}_{out}$ and $ \hat{P}_{\slashed{\partial}}=\hat{I}-\hat{P}_{\partial}$.

The state vector of neurons, ${\bf x} \in  \mathbb{R}^N $, or just state vector, evolves in discrete time-steps according to equation
\be
{\bf x}({t+1}) = {\bf f} \left ( \hat{w} {\bf x}(t)+ {\bf b} \right) \label{eq:discrete}
\ee
which can also be written in terms of components\footnote{Summations over repeated indices are implied everywhere in the paper unless stated otherwise. For example, $w_{ij} x_j = \sum_j w_{ij} x_j$, $\frac{\partial^2 F}{\partial q^2_k} = \sum_k \frac{\partial^2 F}{\partial q^2_k}$ and $\left ( \frac{\partial F}{\partial q_k}\right )^2 = \sum_k \left ( \frac{\partial F}{\partial q_k}\right )^2$.}
\be
{x}_i({t+1}) = f\left ( w_{ij} x_j(t)+ b_i \right).\label{eq:discrete_component}
\ee
 A crucial simplification of the dynamical system \eqref{eq:discrete} was to assume that the activation map ${\bf f}: \mathbb{R}^N \rightarrow \mathbb{R}^N $ acts separately on each component \eqref{eq:discrete_component} with some activation function $f(x)$. Logistic function $f(x) =(1+\exp(x))^{-1}$  and rectified linear unit $f(x) =\max(0,x)$ are some important examples of the activation function, but we shall use the hyperbolic tangent  $f(x) = \tanh(x)$ which is also widely used in machine learning.  The main reason is that the hyperbolic tangent is a smooth odd function with a finite support which greatly simplifies analytical calculations that we shall carry out in the paper.

The main problem in machine learning, or the main learning objective, is to find a bias vector, ${\bf b}$, and a weight matrix, $\hat{w}$, which minimize some suitably defined loss function  ${H}({\bf x}, {\bf b}, \hat{w})$. In what follows we shall consider two loss functions: the ``bulk'' loss and the ``boundary'' loss. The bulk loss function is defined as a local sum over all neurons
\bea
H({\bf x}, {\bf b}, \hat{w}) &=&  \frac{1}{2} \left ( {\bf x}  - {\bf f} \left ( \hat{w} {\bf x}+ {\bf b} \right) \right )^T \left (  {\bf x}  - {\bf f} \left ( \hat{w} {\bf x}+ {\bf b} \right) \right ) +  { V}({\bf x})\notag\\&=&  \frac{1}{2} \left ( {x}_i  - { f} \left ( {w}_{ij} {x_j}+ {b_i} \right) \right ) \left (  {x}_i  - {f}\left ({w}_{ik} {x_k}+ { b_i} \right) \right ) + \sum_i V({x}_i).
\label{eq:bulk_loss} 
\eea
The first term represents the sum over squares of local errors or, equivalently, differences between the state of a neuron before, $x_i$,  and after, $ f\left ( w_{ij} x_j+ b_i \right)$, a single execution of the activation map. The second term represents a local objective such as a binary classification of the signal  $x_i$. For example, if $ V({x}_i) = - \frac{m}{2} {x}_i^2$, then the values of ${x}_i$ closer to lower- and upper-bounds are rewarded and values in-between are penalized.  Although the bulk loss is much easer to analyze analytically, in practice it is often more useful to define the boundary loss function by summing over only boundary neurons,
\be
{H}_\partial({\bf x}, {\bf b}, \hat{w})= {H}(\hat{P}_\partial {\bf x}, \hat{P}_\partial {\bf b}, \hat{P}_\partial^T \hat{w} \hat{P}_\partial).
\ee 
In fact the boundary loss is usually used in supervised learning, but, as was argued in \cite{machine_learning}, the bulk loss is more suitable for unsupervised learning tasks. 

Instead of following the dynamics of the individual states, which might be challenging, one can use the principle of maximum entropy \cite{Jaynes,Jaynes2} to derive a canonical ensemble of states \cite{machine_learning}. The corresponding canonical partition function is 
\be
{\cal Z}(\beta, {\bf b}, \hat{w}) = \int d^{N} x  \;e^{- \beta H\left  ( {\bf x}, {\bf b}, \hat{w} \right )} \label{eq:canonical_ensemble}
\ee
and the free energy is
\be
F(\beta, {\bf b}, \hat{w})  = -\frac{1}{\beta} \log{\cal Z}(\beta, {\bf b}, \hat{w}).
\ee
At a constant ``temperature'', $T=\beta^{-1} = const$, the ensemble can evolve with time either due to  internal (or what we shall call hidden) dynamics of the state vector, ${\bf x}({t})$, or due to the external (or what we shall call training) dynamics of the bias vector, ${\bf b}(t)$, and weight matrix, $\hat{w}(t)$. The partition function for the bulk loss function \eqref{eq:bulk_loss} with a mass-term potential, ${ V}({x}_i) = - \frac{m}{2} {x}_i^2$, and a hyperbolic tangent activation function, $f(x)=\tanh(x)$, was calculated in \cite{machine_learning} using Gaussian approximation. The result is 
\be
{\cal Z}(\beta, {\bf b}, \hat{w}) \approx  (2 \pi)^{N/2} \det \left ( \hat{I} (1-\beta m) +  \beta\hat{G} \right )^{-1/2} \label{eq:Z}
\ee
where
\be
\hat{G} \equiv \left ( \hat{I} - \hat{f}' \hat{w}\right)^T \left ( \hat{I} - \hat{f}' \hat{w}\right)
\ee
and  $\hat{f}'$ is a diagonal matrix of first derivatives of the activation function, 
\be
f'_{ii} \equiv \left (\frac{d f(y_i)}{d y_i}\right )_{y_i = w_{ij} \langle x_j \rangle + b_i}. \label{eq:fd}
\ee

\section{Thermodynamics of learning}\label{sec:Thermodynamics}

Given the partition function, the average loss can be calculated by a simple differentiation,  
\be
 U(\beta , {\bf b}, \hat{w})  = \left \langle H\left  ( {\bf x}, {\bf b}, \hat{w} \right )  \right \rangle = - \frac{\partial}{\partial \beta}  \log({\cal Z}(\beta, {\bf b}, \hat{w})) =  \frac{\partial}{\partial \beta}\left ( \beta F(\beta, {\bf b}, \hat{w}) \right ).
\ee 
If the neural network was trained for a long time, then the weight matrix and the bias vector are in a state which minimizes (at least locally) the average loss function and then its variations with respect to $\hat{w}$ and ${\bf b}$ must vanish, 
 \bea
 \frac{\partial U(\beta , {\bf b}, \hat{w})  }{\partial w_{ij}} &=& \frac{\partial^2}{\partial w_{ij} \partial \beta}  \left (\beta F(\beta, {\bf b}, \hat{w}) \right ) =0 \notag\\
 \frac{\partial U(\beta , {\bf b}, \hat{w})  }{\partial b_{i}} &=& \frac{\partial^2}{\partial b_{i} \partial \beta}   \left (\beta F(\beta, {\bf b}, \hat{w}) \right ) =0. \label{eq:equilibrium}
 \eea
 We shall call this state, the state of the learning equilibrium. An important property of the equilibrium, which follows from \eqref{eq:equilibrium}, is that the total free energy must decompose into a sum of two terms
\be
F(\beta, {\bf b}, \hat{w}) =  A (\beta)  - \frac{1}{\beta}C({\bf b}, \hat{w}). \label{eq:decomposition}
\ee
Likewise, the total entropy must also decompose into a sum of two terms,
\bea
S_x(\beta,{\bf b}, \hat{w} )  &=& \beta^2  \frac{\partial}{\partial \beta } F(\beta, {\bf b}, \hat{w}) =  \beta^2  \frac{\partial}{\partial \beta } \left (  A (\beta)  - \frac{1}{\beta}C({\bf b}, \hat{w}) \right ) ={S}_0(\beta) +  C({\bf b}, \hat{w}) \label{eq:entropy_complexity}
\eea
where the first term is the familiar thermodynamic entropy
\be
{S}_0(\beta) =\beta^2 \frac{\partial A(\beta)}{\partial \beta}= \beta (U (\beta) - A(\beta) ).
\ee
and the second term, $C({\bf b}, \hat{w})$, is related to the complexity of the neural network (see Ref. \cite{machine_learning}). 
  
 As the learning progresses, the average loss, $ U(\beta) $, decreases, the temperature parameter, $\beta^{-1}$, decreases and, thus, one might expect that the thermodynamic entropy, ${S}_0$, should also decrease. However, it is not the thermodynamic entropy, ${S}_0$, but the total entropy, $S_x$, (whose exponent describes accessible volume of the configuration space for ${\bf x}$) should decrease with learning.  We call it the second law of learning:\\
 \\
{\bf Second Law of Learning:} {\it the total entropy of a learning system can never increase during learning and is constant in a learning equilibrium,} 
 \be
 \frac{d}{dt} S_x  \le 0.\label{eq:second_law}
 \ee
 \\
 In the long run the system is expected to approach an equilibrium state with the smallest possible total entropy, ${S}_x$, which corresponds to the lowest possible sum of the thermodynamic entropy, ${S}_0(\beta)$, and of the complexity function $C({\bf b}, \hat{w})$. 
 
For a system transitioning between equilibrium states at constant temperature, $T=1/\beta$, variations of the free energy must vanish, $d F=0$, and  then equation \eqref{eq:decomposition} takes the from of the first law,
\be
 d A  -  T d C = dU - T d S_x  = dU - T d S_0  -   T d C  = 0, \label{eq:first}
 \ee
or what we call the first law of learning:
\\
\\
{\bf First Law of Learning:} {\it the increment in the loss function is proportional to the increment in the thermodynamic entropy  plus the increment   in the complexity} 
\be
dU = T d S_x = T d S_0  +   T d C .\label{eq:first_law}
\ee

\section{Entropic mechanics}\label{sec:Entropic}

So far the neural networks were analyzed by considering statistical ensembles of the state vectors, ${\bf x}$, but the bias vector, ${\bf b}$, and weight matrix, $\hat{w}$, were treated deterministically. The next step is to promote ${\bf b}$ and $\hat{w}$ to stochastic variables in order to study their near-equilibrium dynamics. In the next section we will show that the training dynamics of ${\bf b}$ and $\hat{w}$ can be approximated by Madelung equations with ${\bf x}$ playing the role of the hidden variables. For this reason, we shall refer to the bias vectors and weight matrices as ``trainable'' variables and to the state vectors as ``hidden'' variables. This does not mean that the trainable variables are the quantized versions of the corresponding classical variables, but only that their stochastic evolution near equilibrium can often be described by quantum mechanics.

Consider a family of trainable variables, ${\bf b}({\bf q})$ and $\hat{w}({\bf q})$, parametrized by dynamical parameters $q_{k}$'s where ${k} \in (1,...,K)$. Typically the number of parameters $K$ is much smaller than  $N + N^2$ (i.e. the number of parameters required to describe a generic vector ${\bf b}$ and a generic matrix $\hat{w}$) and the art of designing  a neural architecture is to come up with functions  ${\bf b}({\bf q})$ and $\hat{w}({\bf q})$ which are most efficient in finding solutions. To make the statement more quantitative, consider an ensemble of neural networks described by a probability distribution  ${p}(t, {\bf q})$ which evolves with time according to a Fokker-Planck equation
\bea
\frac{\partial p}{\partial t}&=&   \frac{\partial}{\partial q_{k}}  \left (D  \frac{\partial p}{\partial q_{k}} - \frac{d {q}_k}{dt} p \right) .
\eea
If we assume that the learning evolution (or the drift) is in the direction of the gradient of the free energy,
\be
 \frac{d q_{k}}{d t}  =  \gamma \frac{\partial F}{\partial q_k}  \label{eq:dq}
\ee
then
\be
\frac{\partial p}{\partial t}=  \frac{\partial}{\partial q_{k}}  \left (D  \frac{\partial p}{\partial q_{k}} - \gamma \frac{\partial F}{\partial q_k}  p \right).
 \label{eq:dP}
\ee
This may be a good guess on short-time scales when the free energy does not change much, but in general both ${p}(t, {\bf q})$  and $F(t, {\bf q})$ can depend on time explicitly and implicitly though variable ${\bf q}$. To describe such dynamics we shall employ the principle of stationary entropy production (see Ref. \cite{entropic}): \\ 
\\
{\bf Principle of Stationary Entropy Production}: {\it The path taken by a system is the one for which the entropy production is stationary.}\\
\\
The principle can be thought of as a generalization of both, the maximum entropy principle \cite{Jaynes, Jaynes2} and the minimum entropy production principle \cite{Prigogine,Klein} which is often used in non-equilibrium thermodynamics. In context of neural networks it is beneficial to have large entropy as it implies a higher rate with which new solutions can be discovered. Then the optimal neural architecture should be the one for which the entropy destruction is minimized or, equivalently, the entropy production is maximized. This justifies the use of the principe in context of the optimal learning systems \cite{machine_learning}. 

The Shannon entropy of the distribution ${p}(t, {\bf q})$ (not to confuse with $S_x(\beta,{\bf q} )$) is given by
\be
{S_q }(t) = - \int d^K q \;\;{p}(t, {\bf q}) \log \left ( {p}(t, {\bf q}) \right ). \label{eq:entropy2}
\ee
and using \eqref{eq:dP} the entropy production is given by
\bea
\frac{d S_q }{d t} & = &  - \int d^K q \;{p}  \frac{\partial  \log ({p}) }{\partial t }    - \int d^K q \; \log ({p})  \frac{\partial p}{\partial t }  \notag \\
& =&  - \frac{d}{dt} \int d^K q\; p    - \int d^K q \; \log ({p})  \frac{\partial p }{\partial t }  \notag \\
& =&    - \int d^K q \; \log ({p})   \frac{\partial}{\partial q_{k}}  \left (D  \frac{\partial p}{\partial q_{k}} - \gamma \frac{\partial F}{\partial q_k}  p \right)  \notag
\eea 
which can be simplified (after integrating by parts and ignoring the boundary terms, i.e. by assuming periodic or vanishing boundary conditions), 
\bea
\frac{d S_q }{d t} &=&   \int d^K q \;  \frac{\partial p }{\partial q_{k}}  \left (\frac{D}{p}  \frac{\partial p}{\partial q_{k}} - \gamma \frac{\partial F}{\partial q_k}   \right) \notag\\  &=&   \int d^K q\; \sqrt{p}  \left (- 4 {D} \frac{\partial^2}{\partial q_k^2}  + \gamma \frac{\partial^2}{\partial q_k^2} F   \right)  \sqrt{p}. \label{eq:production}
\eea
This quantity is a functional of both  ${p}(t, {\bf q})$ and  $F(t, {\bf q})$ and,  thus, in addition to modeling the dynamics of the probability distribution we must also model the dynamics of the free energy. 

The total rate of change of the free energy is given by
\be
\frac{d}{dt} F(t, {\bf q})=  \frac{\partial F(t, {\bf q})}{\partial t}  +  \frac{d q_k}{d t} \frac{\partial F(t, {\bf q})}{\partial q_k}  =  \frac{\partial F(t, {\bf q})}{\partial t}  + \gamma  \left ( \frac{\partial F(t, {\bf q})}{\partial q_k} \right )^2 \label{eq:dF}
\ee
where the first term represents the change of the free energy due to dynamics of hidden variables, ${\bf x}$, and the second term represents the change in the free energy due to dynamics of trainable variables, ${\bf b}$  and $\hat{w}$. In what follows, it will be convenient to denote the time-averaged rate of change of free energy as
\be
\left \langle \frac{d}{dt} F(t, {\bf q}) \right \rangle_t \equiv - V({\bf q}). \label{eq:free_energy}
\ee
Then, according to the principle of stationary entropy production, the dynamics of $p(t, {\bf q})$ and $F(t, {\bf q})$ must be such that the entropy production is extremized subject to a constraint 
\be 
\frac{\partial F}{\partial t}  + \gamma  \left ( \frac{\partial F}{\partial q_k} \right )^2 + V  =0.\label{eq:free_energy2}
\ee
The optimization problem can be solved by defining the following ``action'',
\be
{{\cal S}_q}[p, F] =   \int_0^T dt \frac{d S_q }{d t} + \mu \int_0^T dt d^K q  \;p \left (\frac{\partial F}{\partial t}  + \gamma  \left ( \frac{\partial F}{\partial q_k} \right )^2 +  V \right ),\label{eq:Lagrangian}
\ee
where $\mu$ is a Lagrange multiplier, and then the ``equations of motion'' are obtained by setting variations of the action to zero, 
\be
\frac{\delta {{\cal S}_q}}{\delta p} = \frac{\delta {{\cal S}_q}}{\delta F} = 0.\label{eq:eom}
\ee

\section{Quantum mechanics}\label{sec:Quantum}

In the previous section we developed a stochastic description of the trainable variables ${\bf q}$ which describe the weight matrix $\hat{w}({\bf q})$ and the bias vector ${\bf b}({\bf q})$. We argued that on short time-scales the dynamics of the probability distribution ${p}(t, {\bf q})$ and  of the free energy $F(t, {\bf q})$ is given by equations \eqref{eq:dP} and \eqref{eq:dF}, but on longer time-scales an approximate dynamics can be obtained using the principle of stationary entropy production. The corresponding ``action'' is given by  \eqref{eq:Lagrangian} which can be rewritten using \eqref{eq:production},
\be
{{\cal S}_q}[p, F] =   \int_0^T dt\, d^K q  \,\sqrt{p}  \left ( -4 {D} \frac{\partial^2}{\partial q_k^2}  + \gamma \frac{\partial^2}{\partial q_k^2} F  +\mu \frac{\partial F}{\partial t}  + \mu \gamma  \left ( \frac{\partial F}{\partial q_k} \right )^2 + \mu V\right)  \sqrt{p}\label{eq:EM}.
\ee
The five terms on the right hand side represent: 

(1)  $-4 {D}  \frac{ \partial^2}{\partial q_{k}^2}$, entropy production due to stochastic dynamics of $q_k$'s, 

(2) $\gamma \frac{\partial^2 F}{\partial q_k^2}$, entropy production due to learning dynamics of $q_k$'s,   

(3)  $\mu \frac{\partial F}{\partial t}$, free energy production due to dynamics of $x_i$'s

(4)  $\mu \gamma  \left ( \frac{\partial F}{\partial q_k} \right )^2$, free energy production due to learning dynamics of $q_k$'s,
 
 (5) $\mu V$, the (negative of) total time-averaged free energy production.\\
Note that the entropy production due to stochastic dynamics is usually  positive (due to the second law of thermodynamics), but the entropy production due to learning dynamics is usually negative (due to the second law of learning). While the learning entropy production is expected to dominate the dynamics far away from an equilibrium, the stochastic entropy production is expected to give the main contribution near equilibrium.

From \eqref{eq:EM} the equations of motion \eqref{eq:eom} are obtained by setting variations to zero,
\bea
\frac{\delta {{\cal S}_q}[p, F]}{\delta F} &=& \gamma  \frac{\partial^2   }{\partial q_k^2}  p - \mu \frac{\partial }{\partial t} p - 2 \mu \gamma  \frac{\partial }{\partial q_k}  \left ( \frac{\partial F}{\partial q_k} p  \right )   =0 \label{eq:var_F}\\
\frac{\delta {{\cal S}_q}[p, F]}{\delta p} &=&-\frac{4 {D}}{\sqrt{p}}  \frac{ \partial^2 \sqrt{p}}{\partial q_{k}^2} +\gamma \frac{\partial^2 F}{\partial q_k^2}  + \mu \frac{\partial F}{\partial t} + \mu \gamma  \left ( \frac{\partial F}{\partial q_k} \right )^2  + \mu V =0.  \label{eq:var_p}
\eea
It is convenient to define a velocity vector 
\be
u_k \equiv 2 \gamma \frac{\partial}{\partial q_k} F. \label{eq:u}
\ee
and then \eqref{eq:var_F} can be expressed as a  Fokker-Planck equation
\be
\frac{\partial}{\partial t} p = - \frac{\partial}{\partial q_k} \left ( u_k p \right ) +  \boxed{ \frac{\gamma}{\mu} \frac{\partial^2}{\partial q_k^2}  p} \label{eq:neural_FP}
\ee
and  \eqref{eq:var_p} as a  Naiver-Stokes equation (after differentiating with respect to $\frac{\partial}{\partial q_j}$) 
\be
\frac{\partial}{\partial t} u_j +  u_k \frac{\partial}{\partial q_k} u_j +\boxed{ \frac{\gamma}{\mu} \frac{\partial^2}{\partial q_k^2} u_j } = - 2 \gamma \frac{\partial}{\partial q_j} \left ( V  - \frac{4 {D}}{\mu \sqrt{p}}  \frac{ \partial^2 \sqrt{p}}{\partial q_{k}^2} \right ). \label{eq:neural_NS} 
\ee

Several comments are in order. First of all, the  Fokker-Planck equation \eqref{eq:neural_FP} differs from the ``stochastic'' Fokker-Planck equation \eqref{eq:dP}. This is a consequence of our assumption that \eqref{eq:dP} is only valid on very short time scales, while, according to the principle of stationary entropy production, equations \eqref{eq:neural_FP} and  \eqref{eq:neural_NS} must be valid on much longer time-scales. Secondly,  if $\mu>0$ then the kinetic viscosity in the  Naiver-Stokes equation \eqref{eq:neural_NS}, $-\frac{\gamma}{\mu}$,  is negative which is a consequence of the second law of learning. And finally, if we neglect the entropy production due to learning (i.e.  $\gamma \frac{\partial^2 F}{\partial q_k^2}$ in  \eqref{eq:EM}), then the resulting equations of motion  would be the same as \eqref{eq:neural_FP} and  \eqref{eq:neural_NS}, but with terms in boxes set to zero. These are the well known Madelung equations which are equivalent to the Schr\"odinger equation
\be
-  i \sqrt{\frac{4D}{\gamma}}  \frac{\partial }{\partial t } \Psi  =  \left (  4 D \; \frac{ \partial^2 }{\partial q_{k}^2} -  V \right ) \Psi 
\ee
for the wave-function defined as
\be
\Psi \equiv \sqrt{p} \exp \left ( i \sqrt{\frac{\gamma}{4 D}} F\right ). 
\ee
Moreover, in this limit the action \eqref{eq:EM} takes the form of the Schr\"odinger action
\be
 {{\cal S}_q}[\Psi] = \int_0^T dt \int d^K q \;\; \Psi^*   \left (  - 4 D\; \frac{ \partial^2 }{\partial q_{k}^2} + V  - i  \sqrt{\frac{4D}{\gamma}} \frac{\partial }{\partial t} \right )  \Psi .\label{eq:QM}
\ee
Therefore, we conclude that near equilibrium, i.e. when the first term in \eqref{eq:EM} is much larger than the second term, our system can be modeled by quantum mechanics. 

\section{Hamiltonian mechanics}\label{sec:Hamiltonian}

The next step is to consider a non-equilibrium dynamics of the trainable variables which is relevant when the second term in \eqref{eq:EM}  is much larger than the first term. This corresponds to a limit when the entropy destruction is dominated by the learning dynamics and the stochastic entropy production is negligible. The corresponding Fokker-Planck equation remains the same as before \eqref{eq:neural_FP}, but the Naiver-Stokes equation \eqref{eq:neural_NS} is greatly simplified
\be
\frac{\partial}{\partial t} u_j +  u_k \frac{\partial}{\partial q_k} u_j +  \frac{\gamma}{\mu} \frac{\partial^2}{\partial q_k^2} u_j  = - 2 \gamma \frac{\partial}{\partial q_j} V.  \label{eq:neural_NS2} 
\ee
In this limit the dynamics of the free energy $F$ does not depend on the probability distribution $p$ and thus equation \eqref{eq:neural_NS2} decouples from \eqref{eq:neural_FP} and can be solved separately. In terms of the free energy the equation of motion \eqref{eq:var_p} is
\be
- \frac{\partial F}{\partial t} =  V + \gamma  \left ( \frac{\partial F}{\partial q_k} \right )^2 +\frac{\gamma}{\mu} \frac{\partial^2 F}{\partial q_k^2}  \label{eq:neural_NS3}
\ee
which can be though of as a Hamilton-Jacobi equation for the Hamilton's principle function $F$ and a Hamiltonian function
\be
H\left (q_k, \frac{\partial F}{\partial q_k}, \frac{\partial^2 F}{\partial q_k^2}\right ) =  V + \gamma  \left ( \frac{\partial F}{\partial q_k} \right )^2 + \frac{\gamma}{\mu} \frac{\partial^2 F}{\partial q_k^2}. \label{eq:Hamiltonian} 
\ee
Note, however, that in classical mechanics the Hamiltonian function depends only on $q_k$'s and $ \frac{\partial F}{\partial q_k}$'s, but in our case it also depends on one more variable $\sum_k \frac{\partial^2 F}{\partial q_k^2}$. 

From equations \eqref{eq:dq} and \eqref{eq:u} we get
\be
\frac{d q_j}{dt} = \gamma \frac{\partial F}{\partial q_j} = \frac{1}{2} u_j \label{eq:dqt}
\ee
and then \eqref{eq:neural_NS3}  can be rewritten as
\be
\frac{d F}{dt} =  \frac{\partial F}{\partial t} + \frac{d q_k}{dt} \frac{\partial F}{\partial q_k} = - \frac{\gamma}{\mu} \frac{\partial^2}{\partial q_k^2} F   -   V.  \label{eq:F}
\ee
In the limit when the entropy production (due to both learning and stochastic dynamics) is negligible, i.e.  $ \left |  \frac{\gamma}{\mu} \frac{\partial^2}{\partial q_k^2} F   \right | \ll  \left |  V\right |$, equations \eqref{eq:dqt} and  \eqref{eq:F} can be used to obtain classical equations of motion  
\be
\frac{d^2 q_j}{dt^2} =    - \gamma  \frac{\partial V }{\partial q_j}.
\ee
In the opposite limit, $ \left | V \right | \ll  \left | \frac{\gamma}{\mu} \frac{\partial^2}{\partial q_k^2} F  \right |$, the equation for free energy \eqref{eq:F} takes the following form
\be
\frac{\partial F}{\partial t} =    -  \gamma \left ( \frac{\partial F}{\partial q_k} \right )^2 -  \frac{\gamma}{\mu} \frac{\partial^2}{\partial q_k^2} F.
\ee
which has a simple time-independent (i.e. $\frac{\partial F}{\partial t} =0$) solution given by,
\be
F =C_0 +  \frac{1}{\mu} \sum_k \log(C_k + \mu q_k) \label{eq:sol}
\ee
where $C_0$ and $C_k$'s are arbitrary coefficients. Note that $\frac{\partial F}{\partial t} =0$ corresponds to a limit when the change in the free energy production due to dynamics of $x_i$'s is negligible or in other words when the training dataset is not dynamical (as is often the case in machine learning). 

The solution \eqref{eq:sol} has an exact form of the free energy for a canonical ensemble \eqref{eq:Z},
\be
F = \frac{1}{2 \beta}\log \det  ((1-\beta m) + \beta \hat{G})  - \frac{N}{2\beta} \log (2\pi)  = \frac{1}{2 \beta} \sum_i  \log  \left ( (1-\beta m) +  \beta \lambda_i \right ) - \frac{N}{2\beta} \log (2\pi),\label{eq:canonical}
\ee
with $\mu=2\beta$ and the dynamical variables $q_i$ set to the eigenvalues $\lambda_i$ of the operator $\hat{G}$. In this limit the average loss is
\be
U = \frac{\partial (\beta F)}{\partial \beta} = \frac{1}{2}\sum_i \frac{\lambda_i}{1 +  \beta \lambda_i} =    \lambda_i \frac{\partial F}{\partial \lambda_i}, \label{eq:loss}
\ee
where for simplicity we have set the mass parameter to zero, $m=0$. This equation can be thought of as a viral theorem for our learning system where $\frac{\partial F}{\partial \lambda_i} $ is the ``force'' acting on a ``particle'' at position $\lambda_i$.  More generally, the eigenvalues $\lambda_i$'s could be arbitrary functions of $q_i$'s and time $t$ and then 
\bea
 \frac{\gamma}{\mu} \sum_k \frac{\partial^2  F}{\partial q_k^2} &=&  \frac{\gamma}{\mu} \sum_{i,j,k} \frac{\partial^2 F}{\partial \lambda_i \partial \lambda_j} \frac{\partial \lambda_i}{\partial q_k}\frac{\partial \lambda_j}{\partial q_k}  = - \frac{\gamma \beta}{2 \mu}\sum_{i,k} \left ( (1-\beta m) +  \beta \lambda_i \right )^{-2} \left (\frac{\partial \lambda_i}{\partial q_k}\right )^2\notag\\
 &=& -  \frac{2 \gamma \beta}{\mu}\sum_{i,k} \left ( \frac{\partial  F}{\partial \lambda_i} \frac{\partial \lambda_i}{\partial q_k}\right )^2  =-\frac{2 \gamma \beta}{\mu}  \sum_{i,j,k}  \frac{\partial  F}{\partial \lambda_i} \left (   \frac{\partial \lambda_i}{\partial q_k}  \delta_{ij}\frac{\partial \lambda_j}{\partial q_k} \right ) \frac{\partial  F}{\partial \lambda_j}\notag\\
  &=& - \frac{2 \gamma \beta}{\mu}\sum_{i,j,k, m, n} \frac{\partial  F}{\partial q_m} \left (  \frac{\partial q_m}{\partial \lambda_i} \frac{\partial \lambda_i}{\partial q_k}  \delta_{ij}\frac{\partial \lambda_j}{\partial q_k}\frac{\partial q_n}{\partial \lambda_j} \right ) \frac{\partial  F}{\partial q_n}\notag\\  &=& - \frac{2 \gamma \beta}{\mu}\sum_{i,k, m, n} \frac{\partial  F}{\partial q_m} \left (  \frac{\partial q_m}{\partial \lambda_i}   \left ( \frac{\partial \lambda_i}{\partial q_k} \right )^2 \frac{\partial q_n}{\partial \lambda_i} \right ) \frac{\partial  F}{\partial q_n}
 \eea
 where we assumed that $\frac{\partial \lambda_i}{\partial q_j}$  is invertible. This implies that for the canonical free energy \eqref{eq:canonical} the Hamiltonian function \eqref{eq:Hamiltonian}  can be written in terms of only first derivatives of the Hamilton's principle function $F$, 
\be
H\left (q_k, \frac{\partial F}{\partial q_k} \right ) =  V + \gamma \frac{\partial  F}{\partial q_m}\left ( \delta_{mn}  - \frac{2  \beta}{\mu}  \left (  \frac{\partial q_m}{\partial \lambda_i}   \left ( \frac{\partial \lambda_i}{\partial q_k} \right )^2 \frac{\partial q_n}{\partial \lambda_i} \right ) \right ) \frac{\partial  F}{\partial q_n},
\ee
and, thus, the system is Hamiltonian although the kinetic term may not be canonical. 

\section{Hidden variables}\label{sec:Hidden}

We have seen that neural networks can exhibit both quantum (Sec. \ref{sec:Quantum}) and classical (Sec. \ref{sec:Hamiltonian})  behaviors if the dynamics of the trainable variables ${\bf q}$ (or equivalently of the bias vector $\bf b$ and weight matrix $\hat{w}$) is followed explicitly, but the dynamics of the hidden variables (or the state vectors ${\bf x}$) was expressed only implicitly through $\frac{\partial F}{\partial t}$. For this reason it was convenient to think of the state vectors ${\bf x}$ as hidden random variables whose individual dynamics was shadowed by our statistical description. In this section we shall be interested instead in a non-equilibrium dynamics of the hidden variables which is relevant, for example, on the time-scales that are much smaller than thermalization time. 

Recall that the state of the individual neurons evolves according to \eqref{eq:discrete} which can be approximated to the leading order as
\be
\bar{x}^{(0)}_i(t+1)  \approx \left ( \hat{f}_0' \hat{w}\right )_{ij}  \bar{x}^{(0)}_j(t) \label{eq:x0}
\ee
where $\hat{f}'_0 = \hat{f}'$ is the matrix of first derivative of the activation function \eqref{eq:fd}. More generally, we can consider $D$ non-interacting subsystems of states vectors (e.g. $D$ separate sets of training data) denoted by ${\bf x}^{(d)}$ where $d = 1, ..., D$. Then the overall distribution of the state vectors is in general multimodal with $D$ local maxima,  $\bar{\bf x}^{(d)}$, and each of these maxima evolves according to
\be
\bar{x}^{(d)}_i(t+1) \approx \left ( \hat{f}_d' \hat{w}\right )_{ij}  \bar{x}^{(d)}_j(t) \label{eq:xd}
\ee 
where
\be
\bar{x}^{(0)} = \sum_d \bar{x}^{(d)}.
\ee
and
\be
\left (\hat{f}'_d\right)_{ii} \equiv \left (\frac{d f(y_i)}{d y_i}\right )_{y_i = w_{ij} \bar{x}^{(d)}_j  + b_i}. 
\ee
It is convenient to define a continuous time coordinate $\tau$ such that 
\be
\frac{\partial }{\partial \tau}  \bar{x}^{(\mu)}_i(\tau)   = \alpha (\bar{x}^{(\mu)}_i(t+1) - \bar{x}^{(\mu)}_i(t)) \label{eq:dxt}
\ee
where $\mu = 0,1... D$ and $\alpha$ is an auxiliary parameter. Although the different subsystems are represented by different hidden variables ${\bf x}^{(d)}$'s, they are all processed by the very same neural network described by the same trainable variable $\bf b$ and $\hat{w}$. With this respect the hidden variables are not interacting directly with each other, but they are interacting (minimally) through the trainable variables,  $\bf b$ and $\hat{w}$. If such (minimal) interactions are negligible, then $\frac{\partial  \bar{x}^{(c)}_i}{\partial \tau}\frac{\partial  \bar{x}^{(d)}_i}{\partial \tau}\propto \delta_{cd}$ with no summations over index $i$. Then 
\be
 \frac{\partial  \bar{x}^{(0)}_i}{\partial \tau} \frac{\partial  \bar{x}^{(0)}_i}{\partial \tau}  =\sum_{d}  \frac{\partial  \bar{x}^{(d)}_i}{\partial \tau} \frac{\partial  \bar{x}^{(d)}_i}{\partial \tau} \;\;\;\;\;\;\;\;\;\;\text{for all}\;i
 \ee
or
\be
\eta_{\mu\nu} \frac{\partial  \bar{x}^{(\mu)}_i}{\partial \tau} \frac{\partial  \bar{x}^{(\nu)}_i}{\partial \tau}= 0 \;\;\;\;\;\;\;\;\;\;\text{for all}\;i
\label{eq:constraint0}
\ee
where $\eta = \text{diag}(-1,1,...,1)$. However, in general the minimal interactions cannot be ignored and then 
\be
g^i_{\mu\nu} \frac{\partial  \bar{x}^{(\mu)}_i}{\partial \tau} \frac{\partial  \bar{x}^{(\nu)}_i}{\partial \tau}= 0\;\;\;\;\;\;\;\;\;\;\text{for all}\;i
\label{eq:constraint3}
\ee
where the metric tensor $g^i_{\mu\nu}$ describes the strength of the interactions. Of course, such a description is only valid if the minimal interactions are weak which is the assumption we are going to make.

To estimate the dynamics of hidden variables $ \bar{x}^\mu$ we assume that the activation function is linear $\hat{f}_d' = \hat{I}$ (with the slope set to one without loss of generality) and then from \eqref{eq:x0} and \eqref{eq:xd} we have
\be
\bar{x}^{(\mu)}(t+1)   \approx w_{ij} \bar{x}^{(\mu)}_j\label{eq:linear}
\ee
and \eqref{eq:dxt} becomes
\be
\frac{\partial  \bar{x}^{(\mu)}_i}{\partial \tau} \approx \alpha \left ( w_{ij} - \delta_{ij} \right )  \bar{x}_j^{(\mu)}.\label{eq:d_wave}
\ee
According to the second law of learning it is expected that the neural network must have evolved to a network with a very low complexity such as a network whose weight matrix is  a permutation matrix
\be
\hat{w} = \hat{\pi}. \label{eq:permutation}
\ee
For example, consider a permutation matrix with only a single cycle which (up to permutations of elements) is given by
\bea
\pi_{ij} = \begin{cases} 1 \;\;\;\;\text{if} \;i -1 = j\;(\text{mod} \;N)\\
 0 \;\;\;\;\text{otherwise}.
\end{cases}
\eea
Then equation \eqref{eq:d_wave} can be rewritten as
\be
\frac{\partial  \bar{x}^{(\mu)}_i}{\partial \tau} =  \alpha \bar{x}_{i-1 (\text{mod} \;N)}^{(\mu)}(t) - \alpha \bar{x}_i^{(\mu)}(t).  \label{eq:d_wave2}
\ee
If we take a continuous limit by defining $\bar{x}^{(\mu)}(\tau,\sigma)$ such that 
\be
\frac{\partial }{\partial \sigma}  \bar{x}^{(\mu)}(\tau,\sigma)   = \alpha (\bar{x}_i^{(\mu)}(t) -\bar{x}_{i-1 (\text{mod} \;N)}^{(\mu)}(t) )
\ee
then \eqref{eq:d_wave2} becomes
\be
 \frac{\partial  \bar{x}^{(\mu)}}{\partial \tau}  = - \frac{\partial \bar{x}^{(\mu)}}{\partial \sigma}. \label{eq:wavep}
\ee
This equation has a simple solution of a periodic ``right-moving'' wave.  In the light-cone coordinates $\xi^\pm \equiv \tau\pm\sigma$, the equation of motion \eqref{eq:wavep} is
\be
\frac{\partial \bar{x}^{(\mu)}}{\partial \xi^+}  =0 \label{eq:right-wave}
\ee
and the constraint equation \eqref{eq:constraint0} is
\be
\eta_{\mu\nu} \frac{\partial \bar{x}^{(\mu)}  }{\partial \xi^-} \frac{\partial \bar{x}^{(\nu)}}{\partial \xi^-}  =0.\label{eq:right-constraint}
\ee

\section{Relativistic strings}\label{sec:Strings}

In the last section we have shown that an equation for a ``right-moving'' wave \eqref{eq:right-wave} can emerge in a statistical description of $D$ minimally-interacting subsystems of  state vectors.  A natural question arises if a ``left-moving'' wave can also emerge in some limit and if so can the dynamics be described in terms of relativistic strings in an emergent space-time? To answer this question we first note that the permutation weight matrix \eqref{eq:permutation} (with an arbitrary number of cycles) is such that,
\be
\hat{\pi}^T \hat{\pi}  = \hat{\pi} \hat{\pi}^T = \hat{I}
\ee
and thus
\be
\hat{G}\left (\hat{\pi} \right ) =  \left ( \hat{\pi} - \hat{I}\right)^T \left ( \hat{\pi} - \hat{I}\right) = \hat{I}- \hat{\pi} - \hat{\pi}^T + \hat{\pi}^T \hat{\pi}  = \hat{G}\left (\hat{\pi}^T \right ). 
\ee
Since the free energy \eqref{eq:canonical} depends on $\hat{\pi}$ only through $\hat{G}$ the very same ensemble of the state vectors can equally likely evolve either towards $\hat{\pi}$ or towards $\hat{\pi}^T$. However, if the exact state of the microscopic weight matrix is unknown, then one must consider an ensemble which contains both options and then the average state vector is given by 
\be
\bar{ x}^{(\mu)}_i = \frac{1}{2}\int d^N {x}^{(\mu)}   p({\bf x}^{(\mu)}, \hat{\pi}) \; {x}^{(\mu)}_i + \frac{1}{2}\int d^N {x}^{(\mu)}   p({\bf x}^{(\mu)}, \hat{\pi}^T)\;  { x}^{(\mu)}_i = \frac{1}{2}\bar{x}^{(\mu-)}_i + \frac{1}{2} \bar{x}^{(\mu+)}_i \label{eq:right-left}
\ee
where the two terms represent statistical averages with respect to the two distributions. 

Following the analysis of the previous section the dynamics of $\bar{x}^{(\mu-)}_i$ and $\bar{x}^{(\mu+)}_i$ can be obtained from \eqref{eq:d_wave} for the respective weight matrices,  
\bea
 \frac{\partial  \bar{x}^{(\mu-)}_i}{\partial \tau}&\approx&\alpha  \left ( \pi_{ij} - \delta_{ij} \right )  \bar{x}_j^{(\mu-)}\label{eq:d_wave3}\\
 \frac{\partial  \bar{x}^{(\mu+)}_i}{\partial \tau}&\approx& \alpha \left ( \pi^T_{ij} - \delta_{ij} \right )  \bar{x}_j^{(\mu+)} \label{eq:d_wave4}.
\eea
In a continuum limit the equations are given by
\bea
 \frac{\partial  \bar{x}^{(\mu-)}}{\partial \tau}&=& - \frac{\partial \bar{x}^{(\mu-)}}{\partial \sigma}\\
 \frac{\partial  \bar{x}^{(\mu+)}}{\partial \tau}&=& + \frac{\partial \bar{x}^{(\mu+)}}{\partial \sigma}
\eea
whose solutions represent respectively the right- and left-moving waves. Then the dynamics of the hidden variables \eqref{eq:right-left} is indeed given by a $1+1$ dimensional wave equation
\be
\frac{\partial^2 }{\partial \tau^2} \bar{x}^{(\mu)}(\tau,\sigma) = \frac{\partial^2 }{\partial \sigma^2} \bar{x}^{(\mu)}(\tau,\sigma). \label{eq:wave}
\ee
In the light-cone coordinates the wave equation is
\be
\frac{\partial }{\partial \xi^-} \frac{\partial }{\partial \xi^+} \bar{x}^{(\mu)}(\tau,\sigma) =0 \label{eq:wavespm}
\ee
and the constraints
\be
\eta_{\mu\nu} \frac{\partial \bar{x}^{(\mu)}  }{\partial \xi^-} \frac{\partial \bar{x}^{(\nu)}}{\partial \xi^-}  =\eta_{\mu\nu} \frac{\partial \bar{x}^{(\mu)}  }{\partial \xi^+} \frac{\partial \bar{x}^{(\nu)}}{\partial \xi^+} =0.\label{eq:constraintpm}
\ee
The action which gives rise to the wave equations \eqref{eq:wavespm} and constraints \eqref{eq:constraintpm} is the Polyakov action which can be written in a covariant form as
\bea
{\cal A}  &=& \int d\sigma d\tau \sqrt{-h} h^{ab} \eta_{\mu\nu} \frac{ \partial \bar{x}^{(\mu)} }{\partial \xi^a} \frac{ \partial \bar{x}^{(\mu)} }{\partial \xi^b}  \label{eq:string_action}
\eea
where $h_{ab}$ is the world-sheet metric and $h$ is its determinant. 

In summary, we showed that $D$ non-interacting subsystems of the state vectors ${\bf x}^{(d)}$ can be described with $D+1$ scalar fields in $1+1$ dimensions. Alternatively one can view the configuration space of the scalar fields as an emergent space-time and then our system can be described with a motion of relativistic strings in $D+1$ dimensions \eqref{eq:string_action}. This is very similar to what is usually done in string theory, with one major difference. Our strings arise from the dynamics of the average state vectors $\bar{\bf x}^{(\mu)}$ and not from the dynamics of the bias vector ${\bf b}$ and weight matrix $\hat{w}$ which undergo learning. Recall that the trainable variables ${\bf b}$ and $\hat{w}$ (or equivalently ${\bf q}$) near equilibrium can be modeled by quantum mechanics (Sec. \ref{sec:Quantum}) and further away from the equilibrium by classical mechanics (Sec. \ref{sec:Hamiltonian}). In contrast, the state vectors $\bar{\bf x}^{(\mu)}$ represent hidden variables of the quantum theory, but their dynamics (in certain limits) is conveniently described by relativistic strings. 

\section{Emergent gravity}\label{sec:Gravity}

Consider a discrete action for the hidden variables (or state vectors), 
\be
{\cal A}= g^i_{\mu\nu} \left ( \alpha^2 \left \langle {x}^{(\mu)}_i {G}_{ij}  {x}^{(\nu)}_j \right \rangle_x-  \frac{d \bar{x}^{(\mu)}_i }{d \tau}\frac{d \bar{x}^{(\nu)}_i}{d \tau}  \right ). \label{eq:action}
\ee
where $g^i_{\mu\nu}$ describes interactions between the subsystems \eqref{eq:constraint3}. This action is a lot more general than \eqref{eq:string_action}, but it can be approximated by the string action for a flat target space, $g^i_{\mu\nu}=\eta_{\mu\nu}$, for a permutation weight matrix, $\hat{w}=\hat{\pi}$, and for a linear activation function $\hat{f}'_d = \hat{I} $. To study the dynamics in the emergent space-time it is convenient to rewrite \eqref{eq:action} as
\be
{\cal A} = \int d^D X  \sqrt{-g} g_{\mu\nu} {T}^{\mu\nu} \label{eq:action2}
\ee
where $g$ is the determinant of $g_{\mu\nu}$ and
\be
\sqrt{-g} {T}^{\mu\nu} \equiv \left ( \alpha^2  \left \langle {x}^{(\mu)}_i {G}_{ij}  {x}^{(\nu)}_j \right \rangle_x-  \frac{d \bar{x}^{(\mu)}_i }{d \tau}\frac{d \bar{x}^{(\nu)}_i}{d \tau}  \right )\prod_\alpha \delta\left ( X^\alpha -\bar{x}^\alpha_i\right )\label{eq:emtensor}
\ee
is the energy-momentum tensor density. 

The equilibrium dynamics of neural networks was first modeled using the principle of maximum entropy with a constraint imposed on the loss function \cite{machine_learning}, but to study a non-equilibrium dynamics of the trainable variables the principle of the stationary entropy production had to be used with a constraint was imposed on the dynamics of free energy \eqref{eq:free_energy2}. In this section we study a non-equilibrium dynamics of the hidden variables, and so the constraint should be imposed on the action which describes the dynamics of the state vectors \eqref{eq:action2}.  Then, according to the principle of stationary entropy production, the quantity which must be extremized is
\be
{\cal S}_x[g] =\int d^{D+1} X \sqrt{-g}  {\cal R}(g) +  \kappa \left ( A - \int d^{D+1} X \sqrt{-g} g_{\mu\nu}    {T}^{\mu\nu} \right ) \label{eq:Lx}
\ee
where $\sqrt{-g}{\cal R}(g)$ is the local entropy production density, $ \kappa$ is a Lagrange multiplier and $A$ is a constant which represents average ${\cal A}$. Note that the energy momentum tensor density \eqref{eq:emtensor} does not depend on the metric  and so varying the corresponding term in \eqref{eq:Lx} with respect to the metric produces the desired result
\be
\frac{\delta}{\delta g_{\alpha\beta}} \left ( \int d^{D+1} X \sqrt{-g} g_{\mu\nu}  {T}^{\mu\nu} \right ) =  \sqrt{-g}  {T}^{\alpha\beta}.
\ee
However, if we are not following the microscopic dynamics of all of the elements of the bias vector and weight matrix, then it is more useful to define 
\be
{\cal L}_M(g, Q)  \equiv  -  g_{\mu\nu}   \left \langle  {T}^{\mu\nu} \right \rangle_{q}
\ee
where $Q$ represents the trainable variables in ${\bf q}$  (or equivalently in ${\bf b}$ and $\hat{w}$) which were not averaged over. Then the action \eqref{eq:Lx} can be written as
\be
{\cal S}_x[g,  Q] = \int d^{D+1} X \sqrt{-g} \left ({\cal R}(g)  + \kappa {\cal L}_M(g, {Q}) \right ) + \kappa {A} \label{eq:GR}
\ee
where ${\cal L}_M(g, Q)$ plays the role of the ``matter'' Lagrangian and then the energy momentum tensor should be defined as
\be
\sqrt{-g}  {\cal T}^{\alpha\beta} \equiv  - \frac{\delta}{\delta g_{\alpha\beta}} \left ( \int d^{D+1} X \sqrt{-g} {\cal L}_M(g, Q)  \right ).
\ee
The parameter $\kappa$ is a Lagrange multiplier which imposes a ``global'' constraint 
\be
\frac{\delta {\cal S}_x[g] }{\delta \kappa} =  {A} +  \int d^{D+1} X \sqrt{-g} {\cal L}_M(g, {Q}) = 0 \label{eq:A}
\ee
but one can also impose the constraint ``locally'' by demanding that 
\be
{A} =  \frac{2}{\kappa} \int d^{D+1} X \sqrt{-g} \Lambda 
\ee
and then the total action becomes
\be
{\cal S}_x[g,  Q] = \int d^{D+1} X \sqrt{-g} \left (  {\cal R}(g) - 2 \Lambda + \kappa {\cal L}_M(g, {Q}) \right ) \label{eq:GR}
\ee
where $ \Lambda $ is the ``cosmological constant''.

Recall that the deviations of the metric $g_{\mu\nu}({\bf X})$ (or $g^i_{\mu\nu}$) from the flat metric $\eta_{\mu\nu}$ represent local interactions between subsystems \eqref{eq:constraint3}. Therefore, if our system is in the process of equilibration, then the entropy production should be a local function of the metric tensor. Using a phenomenological approach due to Onsager \cite{Onsager} we can expand the entropy production around equilibrium \cite{emergentgravity}, 
\be
\sqrt{-g} {\cal R} = \sqrt{-g}  L^{\mu\nu\;\alpha\beta\;\gamma\delta} g_{\alpha\beta, \mu}  g_{\gamma\delta, \nu } \label{eq:Onsager2}.
\ee
where
\be
g_{\alpha\beta, \mu } \equiv \frac{\partial g_{\alpha\beta} }{\partial X^{\mu }}
\ee
and $\sqrt{-g}  L^{\mu\nu\;\alpha\beta\;\gamma\delta}$ is the Onsager tensor density. The overall space of such tensors is pretty large, but it turns out that a very simple and highly symmetric choice leads to general relativity:
\be
\sqrt{-g} L^{\mu\nu\;\alpha\beta\;\gamma\delta} =  \frac{1}{4} \sqrt{-g}  \left ( 2 g^{\alpha\gamma} g^{\beta\nu}g^{\mu\delta} - g^{\alpha\gamma} g^{\beta\delta} g^{\mu\nu} - g^{\alpha\beta} g^{\gamma\delta} g^{\mu\nu} \right ) .\label{eq:Onsager}
\ee
After integrating by parts, neglecting boundary terms and collecting all other terms we get
\bea
\int d^{D+1} X &\sqrt{-g} &{\cal R}  =
  \int d^{D+1} X \sqrt{-g} g^{\mu\nu} \;\; 2\left (\Gamma^{\alpha}_{\phantom{\alpha}\nu[\mu, \alpha]} +  \Gamma^{\beta}_{\phantom{\beta}\nu[\mu} \Gamma^\alpha_{\phantom{\alpha}\alpha]\beta}\right )=\\\notag  
  &=&\int d^{D+1} X \sqrt{-g} \;\; \frac{1}{4} \left (  2 g^{\alpha\gamma} g^{\beta\nu} g^{\mu\delta} -g^{\alpha\gamma} g^{\beta\delta} g^{\mu\nu} -g^{\alpha\beta} g^{\gamma\delta} g^{\mu\nu} \right )  g_{\alpha\beta,\mu} g_{\gamma\delta,\nu}\;
\eea
 where
 \be
 \Gamma^{\mu}_{\phantom{\mu}\gamma\delta} \equiv \frac{1}{2} g^{\mu\nu} \left ( g_{\nu\gamma,\delta} +  g_{\nu\delta,\gamma} -  g_{\gamma\delta, \nu} \right ) 
 \ee
 and
 \be
  \Gamma^\alpha_{\phantom{\alpha}\mu\nu, \beta} \equiv \frac{\partial}{\partial X^\beta}  \Gamma^\alpha_{\phantom{\alpha}\mu\nu}.
 \ee
Thus, upon varying \eqref{eq:Lx} with respect to the metric we get the Einstein equations
\be
{\cal R}_{\mu\nu} - \frac{1}{2} {\cal R}g_{\mu\nu} + \Lambda g_{\mu\nu} = \kappa  {\cal T}_{\mu\nu}
\ee
where the Ricci tensor is defined as ussual
\be
{\cal R}_{\mu\nu} \equiv 2 \left (\Gamma^{\alpha}_{\phantom{\alpha}\nu[\mu, \alpha]} +  \Gamma^{\beta}_{\phantom{\beta}\nu[\mu} \Gamma^\alpha_{\phantom{\alpha}\alpha]\beta}\right ).
\ee
Note that according to definition \eqref{eq:Onsager} the Onsager tensor need not be positive definite which would be inconsistent with the second law of thermodynamics, but is permitted by the second law of learning. 

\section{Holography}\label{sec:Holography}

In the preceding sections we applied the principle of the stationary entropy production to study the dynamics of the neural networks in two different limits. In the first limit the trainable variables ${\bf q}$ were treated stochastically, but their dynamics was constrained by the hidden variables ${\bf x}$ through the free energy, $F$. The resulting dynamics of the system was shown to exhibit quantum and classical behaviors described by the functional ${\cal S}_q[p,F]$ (see \eqref{eq:EM}). In the second limit the hidden variables  ${\bf x}$ were treated stochastically, but their dynamics was constrained by the trainable variables ${\bf q}$ through the action, ${\cal A}$. The resulting dynamics of the system was shown to exhibit a behavior described by the action of a gravitational metric theory, such as general relativity,  ${\cal S}_x[g, Q]$ (see \eqref{eq:GR}). The two limits are certainly very different: the ``gravitational'' theory describes very sparse and deep neural networks and in the ``quantum'' theory the network can be very dense and shallow. However, one might wonder if it may possible to map the sparse and deep neural network to the dense and shallow neural network without losing the ability of the neural network to learn. If the answer is affirmative, then this would imply that the two descriptions - quantum and gravitational (or dense and sparse, or  shallow and deep) - are dual and either one can be used to describe the learning dynamics. 

In this section we shall explore an idea that the duality not only exists, but is also holographic in a sense that the degrees of freedom of the gravitational theory, i.e. ${\bf x}$, ${\bf b}$ and $\hat{w}$, can be mapped to only boundary degrees of freedom of the quantum theory, i.e. ${\bf x}^\partial$, ${\bf b}^\partial$ and $\hat{w}^\partial$. The non-equilibrium dynamics of both systems is governed by the principle of stationary entropy production and to justify such a mapping the entropy production of the gravitational system $\Delta S_x$ should correspond to the entropy production of the quantum system $\Delta S^\partial_q$.  Roughly speaking, this means that the uncertainty in the position of neurons in the bulk, ${\bf x}$, should correspond to the uncertainty in the values of quantum variables on the boundary, i.e. ${\bf b}^\partial$ and $\hat{w}^\partial$. For example, consider a mapping defined by 
\bea
{\bf x}^\partial &=& \hat{P}_\partial {\bf x}\\
{\bf b}^\partial &=& \hat{P}_\partial  {\bf b}\\
\hat{w}^\partial(\epsilon) &=& \hat{P}_\partial \frac{\epsilon \hat{w} }{\hat{I} -\epsilon \hat{w}} \hat{P}^T_\partial
\eea
In a microscopic picture the gravitational system consists of long chains of neurons (see Sec. \ref{sec:Hidden}) connecting different pairs of the boundary neurons, $i$ and $j$, but the length of these chains is encoded in the elements of the  boundary weight matrix,
\be
d(i, j) = \log_\epsilon \left ({w}^{\partial}_{ij}(\epsilon) \right ) - 1.
\ee
The smaller the element ${w}^{\partial}_{ij}$, the larger the number of intermediate bulk neurons connecting $i$ to $j$. Whenever any two chains of neurons $i$-$j$ and $k$-$l$ have a chance of intersecting and forming two other chains of neurons $i$-$l$ and $k$-$j$, the entropy of the bulk theory changes. On the other side of the duality,  the same event can lead to the corresponding elements $w^\partial_{ij}$, $w^\partial_{kl}$, $w^\partial_{kj}$ and $w^\partial_{il}$ to change or, in other words, to the entropy production in the boundary theory. Thus, it is not too unreasonable to expect that the entropy production in both system are related.

The holographic duality can be formulated more precisely by considering the action functionals which determine the dynamics in both theories. In the boundary theory the  action ${\cal S}_q\left [p({\bf q}^\partial),F({\bf q}^\partial ) \right]$ is given by equation \eqref{eq:EM} and in the bulk theory the action ${\cal  S}_x[g({\bf X}),  Q({\bf X})]$ is given by equation   \eqref{eq:GR}. For the two systems to be dual the two actions must be proportional
\be
 {\cal  S}_x[g({\bf X}), Q({\bf X})] \sim {\cal S}_q\left [p({\bf q}^\partial ),F({\bf q}^\partial) \right],\label{eq:actions}
\ee
or, using  \eqref{eq:EM} and \eqref{eq:GR},
\bea
 \int d^{D+1} X \sqrt{-g} && \left ( {\cal R}(g) - 2 \Lambda + \kappa{\cal L}_M(g, {Q})  \right )\label{eq:duality} \\ \notag
 &\sim& \int dt\, d^{K^\partial}q^\partial  \,\sqrt{p}   \left ( -4 {D} \frac{\partial^2}{\partial q_k^2}  + \boxed{\gamma\left ( \frac{\partial}{\partial q^\partial_k}\right )^2  F } +\mu \frac{\partial F}{\partial t}  + \mu \gamma  \left ( \frac{\partial}{\partial q^\partial_k}\right )^2 F + \mu V\right)  \sqrt{p}.
\eea
The left hand side describes the bulk gravitational theory, the right hand side describes the boundary theory and the duality transformation is nothing but changes of variables between $\left ( g,Q\right )$ and $\left (p, F\right )$.  Note, however, that the boundary theory can be approximated by quantum mechanics only in the limit when the entropy production due to learning (i.e. the quantity in the box in \eqref{eq:duality}) is subdominant. Therefore the holography described by \eqref{eq:actions} should be considered as more general than the holography discussed, for example, in context of the AdS/CFT correspondence where the CFT side is quantum and the AdS side is gravitational.

\section{Discussion}\label{sec:Discussion}

In this paper we discussed a possibility that the entire universe on its most fundamental level is a neural network. This is a very bold claim. We are not just saying that the artificial neural networks can be useful for analyzing physical systems \cite{Carleo:2019ptp} or for discovering physical laws \cite{Tegmark}, we are saying that this is how the world around us actually works. With this respect it could be considered as a proposal for the theory of everything, and as such it should be easy to prove it wrong. All that is needed is to find a physical phenomenon which cannot be described by neural networks. Unfortunately (or fortunately) it is easer said than done. It turns out that the dynamics of neural networks is so complex that one can only understand it in very specific limits. The main objective of this paper was to describe the behavior of the neural networks in the limits when the relevant degrees of freedom  (such as bias vector, weight matrix, state vector of neurons) can be modeled as stochastic variables which undergo a learning evolution. In this section we shall briefly discuss the main results and implications of the results for a possible emergence of quantum mechanics, general relativity and macroscopic observers from a microscopic neural network. 

Emergent quantum mechanics is a relatively new \cite{Adler, tHooft}, but rapidly evolving field \cite{Blasone,Grossing,Acosta,deCordoba, Caticha, entropic}  which is based on a set of very old ideas, dating back to the works of de Brogie and Bohm. The de Broglie-Bohm theory (also known as pilot wave theory or Bohmian mechanics) was originally formulated in terms of  non-local hidden variables \cite{Bell} which makes it an easy target. The main new insight is that quantum mechanics may not be a fundamental theory, but only a mathematical tool which allows us to carry out statistical calculations in certain dynamical systems. If correct, then one should be able to derive all of the essential ingredients (complex wave-function, Schr\"odinger equation, etc.) from first principle. In this paper we did exactly that for a dynamical system of a neural network which contains two different types of degrees of freedom: trainable (e.g. bias vector and weight matrix) and hidden (e.g. state vector of neurons). What we showed is that the dynamics of the trainable variables near equilibrium is described by Madelung (or equivalently Schr\"odinger) equations with free energy (for a canonical ensemble of hidden variables) representing the quantum phase (see Sec. \ref{sec:Quantum}), and further away from the equilibrium their dynamics is described by Hamilton-Jacobi equations with free energy representing the Hamilton's principal function (see Sec. \ref{sec:Hamiltonian}). This demonstrates that the neural networks can indeed exhibit emergent quantum and also classical behaviors. It is important to emphasize that the learning dynamics was essential and the stochastic dynamics alone would not have produced the desired result. 

Emergent (or entropic) gravity is also a relatively new field \cite{Jacobson,Verlinde, Padmanabhan}, but it is far less clear if or when progress is being made. The main problem is that emergent gravity is not just about gravity, but is also about emergent space \cite{Swingle, Dong, Carroll, graphflow}, emergent Lorentz invariance \cite{Laughlin, Sibiryakov, spacetime},  emergent general relativity \cite{Barcelo, Cao, emergentgravity} etc. Quite remarkably, neural networks open up a new avenue to address all these problems in context of the learning dynamics. It turns out that a dynamical space-time can indeed emerge from a non-equilibrium evolution of the hidden variables (i.e. state vector of neurons) in a manner very similar to string theory. In particular, if one considers $D$ minimally-interacting (trough bias vector and weight matrix) subsystems with average state vectors, $\bar{\bf x}^{1}$, ..., $\bar{\bf x}^{D}$ (and the total average state vector $\bar{\bf x}^{0}$) then the dynamics of $\bar{\bf x}^{\mu}$ can be modeled with relativistic strings in an emergent $D+1$ dimensional space-time (see Secs.  \ref{sec:Hidden} and \ref{sec:Strings}) and if the interactions are described by a metric tensor, then the dynamics can be modeled with Einstein equations (see Sec. \ref{sec:Gravity}). Once again, not only stochastic, but also learning dynamics was essential for the equilibration of the emergent space-time to exhibit behavior of a gravitational theory such as general relativity.  This demonstrates that the dynamics of a neural network in the appropriate limits can be approximated by both emergent quantum mechanics and emergent general relativity, but the two limits are very different. The gravitational theory describes very sparse and deep neural networks and in the quantum theory the neural network can be very dense and shallow. However, it is possible that there exists a holographic duality map between the bulk neurons of the deep and sparse network to the boundary neurons of the shallow and dense network (see Sec. \ref{sec:Holography}). 

We now come to one of the most controversial questions: how can macroscopic observes emerge in a physical system? The question is extremely important not only for settling some philosophical debates, but for understanding the results of real physical experiments \cite{Bell} and cosmological observations \cite{Measure}. As was already mentioned, our current understanding of fundamental physics does not allow us to formulate a self-consistent and paradoxes-free definition of observers and a possibility that observers is an emergent phenomenon is certainly worth considering. Indeed, if both quantum mechanics and general relativity are not fundamental, but emergent phenomena, then why cannot macroscopic observers also emerge in some way from a microscopic neural network. Of course this is a lot more difficult task and we are not going to resolve it completely, but we shall mention an old idea that might be relevant here. It is the principle of natural selection.  We are not talking about cosmological natural selection \cite{Smolin2}, but about the good old biological natural selection \cite{Darwin}, although the two might actually be related. Indeed, if the entire universe is a neural network, then something like natural selection might be happening on all scales from cosmological ($>10^{+15}$ m) and biological ($10^{+2} - 10^{-6}$ m) all the way to subatomic ($<10^{-15}$ m) scales. The main idea is that some local structures (or architectures) of neural networks are more stable against external perturbations (i.e. interactions with the rest of the network) than other local structures. As a result the more stable structures are more likely to survive and the less stable structures are more likely to be exterminated.  There is no reason to expect that this process might stop at a fixed time or might be confined to a fixed scale and so the evolution must continue indefinitely and on all scales. We have already seen that on the smallest scales the learning evolution is likely to produce structures of a very low complexity (i.e. second law of learning) such as one dimensional chains of neurons, but this might just be the beginning. As the learning progresses these chains can chop off loops, form junctions and according to natural selection the more stable structures would survive. If correct, then what we now call atoms and particles might actually be the outcomes of a long evolution starting from some very low complexity structures and what we now call macroscopic  observers and biological cells might be the outcome of an even longer evolution. Of course, at present the claim that natural selection may be relevant on all scales is very speculative, but it seems that neural networks do offer an interesting new perspective on the problem of observers.

{\it Acknowledgments.}  This work was supported in part by the Foundational Questions Institute (FQXi).

\end{document}